\ifdefined\ReviewLayout
  \documentclass[preprint,12pt]{elsarticle}
\else
  \documentclass[final,5p,twocolumn]{elsarticle}
\fi
\usepackage[T1]{fontenc}
\usepackage{amsmath,amsthm,mathtools}
\usepackage{newtxtext,newtxmath}
\usepackage{bm}
\usepackage{graphicx,booktabs,tabularx,array}
\usepackage{microtype,xurl}
\usepackage{tikz,tikz-cd}
\usetikzlibrary{arrows.meta,positioning,calc}
\usepackage{listings}
\usepackage[hidelinks]{hyperref}
\hypersetup{pdftitle={Typed Transverse Momentum Parton distribution function computations},pdfauthor={Dustin Keller}}
\journal{Computer Physics Communications}
\biboptions{sort&compress}
\newtheorem{theorem}{Theorem}
\newtheorem{lemma}[theorem]{Lemma}

\newcommand{\GeV}{\,\mathrm{GeV}}
\newcommand{\dd}{\mathrm d}
\newcommand{\Eval}{\operatorname{Eval}}
\newcommand{\Dep}{\operatorname{Dep}}
\newcommand{\erase}{\operatorname{erase}}
\newcommand{\safe}{\operatorname{safe}}
\newcommand{\PlanCheck}{\operatorname{check}}
\newcommand{\cacheable}{\operatorname{cacheable}}
\newcommand{\Var}{\mathsf{Var}}
\newcommand{\Atom}{\mathsf{Atom}}
\newcommand{\Unary}{\mathsf{Unary}}
\newcommand{\Binary}{\mathsf{Binary}}
\newcommand{\Live}{\mathsf{Live}}
\newcommand{\Cache}{\mathsf{Cache}}
\newcommand{\den}[1]{\llbracket #1\rrbracket}
\newcommand{\softcode}[1]{\texttt{\small #1}}
\newcolumntype{Y}{>{\raggedright\arraybackslash}X}
\begin{document}
\raggedbottom
\begin{frontmatter}
\title{Typed Transverse Momentum Parton distribution function computations}
\author[uva]{Dustin Keller}
\affiliation[uva]{organization={Department of Physics, University of Virginia},
 city={Charlottesville},postcode={22904},state={Virginia},country={USA}}
\begin{abstract}
A transverse-momentum-dependent parton distribution enters an extraction through a specified observation map and a parameter-dependent numerical calculation. Equal units and matching array dimensions do not establish that two implementations represent the same inference problem. We present typed-TMD, a computational framework that makes observation semantics, parameter dependencies, and admissible evaluation plans explicit. In a controlled study using a pinned arTeMiDe/SV19 Drell--Yan calculation, replacing finite-bin averages by transverse-momentum midpoint predictions changes the noiseless fitted evolution parameter from \(0.030\) to \(0.073\;\mathrm{GeV}^2\), while the incorrect fit retains a modest minimum $\chi^2$ of $1.56$. The observation contract rejects this composition before fitting. At a separate computational layer, a concrete finite-expression checker is proved sound in Lean: accepted dependency-safe caching preserves the complete parameterized function for every admitted parameter variation under a fixed pure interpretation. A category of variation-aware interfaces establishes when independently checked stages compose, including transport of cache reference states. Literal certificates instantiate the result on selected complete plans from the numerical calculations. Independent observation quadrature, controlled closure and coverage diagnostics, and same-backend timing measurements delimit the numerical and statistical evidence. The framework preserves scientific meaning through supported computational transformations while keeping physical assumptions, exact finite-plan proofs, and numerical validation distinct. The implementation, proof sources, examples, and reproduction material are released as version 0.4.0.
\end{abstract}
\begin{keyword}
transverse-momentum-dependent distributions \sep Drell--Yan \sep scientific software \sep formal verification \sep Lean \sep category theory \sep reproducibility
\end{keyword}
\end{frontmatter}

\section*{Program summary}
\noindent\textit{Program title:} typed-TMD.\\
\textit{Version:} 0.4.0.\\
\textit{Persistent identifier:} \href{https://doi.org/10.5281/zenodo.23546800}{10.5281/zenodo.23546800}.\\
\textit{Repository:} \url{https://github.com/zetanaut/typed-TMD}.\\
\textit{Licensing provisions:} MIT for original code; separately obtained upstream software and inputs retain their own terms.\\
\textit{Programming languages:} Python and Lean 4.21.0; the optional external backend uses Fortran, f2py, and LHAPDF C++.\\
\textit{Requirements:} Python 3.11 or a compatible tested environment, NumPy, and SciPy for numerical replay. Formal replay additionally requires the pinned Lean toolchain; no mathlib dependency is used. Native Drell--Yan reproduction requires the separately pinned arTeMiDe, LHAPDF, and PDF inputs.\\
\textit{Tested platforms:} macOS/arm64 and Linux for the declared runtime and replay scopes; native backend and kernel-certificate executions are identified separately in the evidence.\\
\textit{Nature of problem:} Interface-compatible numerical arrays may represent different observations, and caching a fitted dependency can change the function being optimized while preserving reference-point values.\\
\textit{Solution method:} Immutable scientific contexts, checked observation operators, explicit finite expressions and variation masks, a Lean-proved cache validator, and compositional reference transport.\\
\textit{Restrictions:} One audited external process/interface and a finite-expression caching language. Numerical accuracy, native primitive semantics, and physical factorization remain separately assessed.\\
\textit{Representative running time:} About 217 s for four five-start, one-parameter external fits in the recorded Mac run, excluding startup and mean generation. Formal checking is a separate one-time workload. Peak memory was not measured.

\section{Introduction}
\label{sec:introduction}
A fitted transverse-momentum-dependent parton distribution (TMD) is meaningful together with its operator convention, perturbative ingredients, collinear inputs, evolution prescription, and use in an observable. Comparisons between factorization formulations already require transformations of nonperturbative and perturbative quantities together~\cite{CollinsRogers2017}. At the implementation level, a further condition is equally important: the program must preserve the observation and the parameterized calculation that the user intended to fit.

Consider two arrays with units of picobarns per GeV. One may contain averages over finite transverse-momentum bins; the other may contain differential predictions at their midpoints. Both are well-formed numerical inputs to a least-squares routine, but they are different observation operators. An optimizer may compensate for their difference by changing a nonperturbative parameter. Agreement between two forward routines at isolated points, or a moderate final fit residual, need not expose that mistake. A separate failure can occur during optimization of the numerical program itself: caching an evolution factor at a reference parameter may leave all reference-point predictions unchanged while removing a fitted sensitivity.

This work addresses these failures at two connected layers. First, a runtime comparison contract records the process and observation semantics, the source of numerical inputs, and the statistical problem. It admits or rejects supported combinations before expensive evaluation. Second, a concrete finite-expression language makes the parameter dependence of evaluation plans explicit. A Lean-verified checker establishes that a supported caching transformation preserves the entire function on its declared variation domain. A category of parameter interfaces describes how such constructions compose and determines the reference state required by a subsequent cache.

\begin{figure*}[t]
\centering
\resizebox{.98\textwidth}{!}{%
\begin{tikzpicture}[>=Latex,font=\small,
 every node/.style={align=center}, core/.style={text width=2.9cm}]
\node[core] (context) at (0,0) {Physical inputs\\and observation};
\node[core] (expr) at (4.5,0) {Finite expression\\$e(\theta)$};
\node[core] (plan) at (9,0) {Candidate plan\\$p(\theta_0,\theta)$};
\node[core] (out) at (13.5,0) {Prediction vector\\and inference};
\draw[->] (context.east) -- (expr.west);
\draw[->] (expr.east) -- (plan.west);
\draw[->] (plan.east) -- (out.west);
\node[font=\scriptsize,text width=2.4cm] at (2.25,.8) {audited\\lowering};
\node[font=\scriptsize,text width=2.7cm] at (6.75,.8) {dependency-safe\\proposal};
\node[font=\scriptsize,text width=2.6cm] at (11.25,.8) {same accepted\\plan};
\node[core] (contract) at (0,-1.4) {Runtime compatibility\\and source binding};
\node[text width=3.5cm] (check) at (6.75,-1.4) {Literal Lean certificate\\erasure + cache safety};
\node[core] (num) at (13.5,-1.4) {Numerical checks\\and refinement};
\draw[->] (contract.north) -- (context.south);
\draw[->] (expr.south) -- (check.north west);
\draw[->] (plan.south) -- (check.north east);
\draw[->] (num.north) -- (out.south);
\node[font=\scriptsize] at (6.75,-2.4) {Exact theorem: $\Eval_M(p;\theta_0,\theta)=\den{e}_M(\theta)$ for every admitted $\theta$.};
\end{tikzpicture}}
\caption{The three evidence layers used in this work. Scientific lowering and observation semantics are supported by runtime contracts and source-level validation. Lean proves preservation of the supported finite representation under caching. Separate numerical comparisons assess the realization of its primitive inputs and evaluations. Neither the schematic arrows nor a successful cache certificate assert an end-to-end proof of the external physics engine. The explicit TikZ source was prepared with AI-assisted drafting and checked against the described workflow.}
\label{fig:workflow}
\end{figure*}
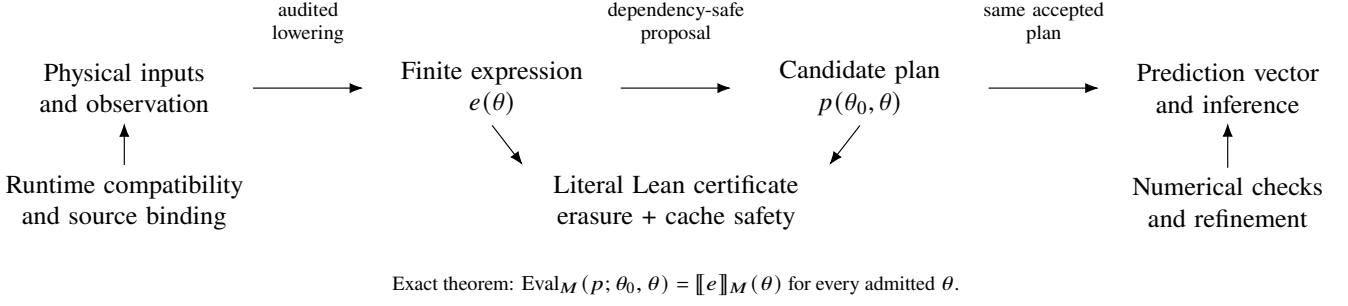

The distinction between the layers is essential. A proof about a finite numerical plan does not establish that its source formula is the correct observable. Conversely, a correctly specified observation does not justify an implementation optimization that changes a fitted dependency. Figure~\ref{fig:workflow} separates scientific assembly, exact finite-plan preservation, and numerical execution. The implementation and evidence are available in typed-TMD v0.4.0~\cite{TypedTMD2026}.

Existing libraries provide the input and calculation infrastructure used in TMD phenomenology. LHAPDF supplies a common interface to collinear parton distributions~\cite{LHAPDF6}; TMDlib2 supplies TMD parameterizations and associated plotting tools~\cite{TMDlib2}; NeoPDF supplies interpolation and distribution classification~\cite{NeoPDF}. Numerical Fourier transformation is itself a significant component of TMD computations~\cite{KangFourier}. arTeMiDe implements TMD predictions and phenomenological ingredients used in global analyses~\cite{ScimemiVladimirov2020,Artemide205}. The present layer neither replaces these calculations nor asserts that conventional software cannot perform its checks. Its contribution is to make comparison conditions, transformations, and evidence reusable parts of the calculation rather than conventions reconstructed independently for each analysis.

The formal ingredients are likewise established. Free term algebras, substitution, and categorical interpretation organize the representation~\cite{Riehl2016}. Semantic preservation is a familiar correctness objective for translation and compilation~\cite{Leroy2009}, and Lean provides a language and proof kernel for checking the required statements~\cite{Lean4}. HepLean illustrates the use of formal definitions and proofs in high-energy physics~\cite{HepLean}. Here the result is tied to an implemented finite-plan checker and literal numerical plans, with explicit boundaries between proof, interpretation, and execution.

The principal application uses seven published ATLAS Drell--Yan bins evaluated with a pinned arTeMiDe/SV19 calculation. A controlled point-versus-bin misuse is rejected before fitting; bypassing the contract in an isolated test measures its inference consequence. A second example freezes a fitted evolution factor, demonstrating a reference-point agreement that does not preserve the parameterized function. The formal checker excludes that transformation. A transparent two-parameter model supplies supporting closure, coverage, and representation tests. These examples have different roles and accuracy prescriptions; their evidence is reported separately.

\section{Scientific contexts and observation contracts}
\label{sec:contracts}
\subsection{A calculation carries more than its values}
A scientific context $\Gamma$ records the operator-sector and process declarations, conventions and units, coefficient inventory, collinear-input identities, kinematic domain, parameter family, and observation definition. The inference context additionally records the ordered data, covariance, normalization treatment, and admissible parameter domain. Immutable records prevent a previously assembled object from changing its declared interpretation unnoticed.

An operation identifies its input and output interfaces. An evolution map, for example, carries scale endpoints and a generator identity; composition requires a common intermediate endpoint and compatible contexts. Replacing a reference boundary uses a distinct transport operation, retaining the original family rather than silently assigning a new ansatz. A prediction is also tagged by its observation operation, not just its array shape. Table~\ref{tab:contracts} summarizes the distinctions exercised here.

\begin{table}[t]
\caption{Implemented contracts and their scientific roles. Runtime compatibility is distinct from a proof that a supplied physical declaration is true.}
\label{tab:contracts}
\small
\begin{tabularx}{\columnwidth}{@{}p{.29\columnwidth}Y@{}}
\toprule
Operation & Required information and outcome\\
\midrule
Observation assembly & Density measure, point/bin operation, boundaries, cuts, normalization, and row order; reject an unmatched observation.\\
Model evaluation & Full parameter state, provider/model identity, domain, and primitive binding; reject a changed context or illegal fixed-coordinate variation.\\
Cache preparation & Computed expression dependencies and variation mask; reject a fitted dependency before initialization.\\
Staged composition & Admitted source/target interfaces and transported reference environment; preserve the fixed-coordinate slice.\\
Fit construction & Matching observation and statistical contexts; admit a model change only under an explicit alternative-model declaration.\\
\bottomrule
\end{tabularx}
\end{table}

Python annotations document the interfaces; validated constructors and runtime checks implement the scientific restrictions. These restrictions are not a complete dependent type system for QCD. In particular, a string naming a subtraction convention does not prove factorization, and a hash identifies bytes rather than validating their physics. Source inspection, independent comparisons, and numerical tests support the interpretation of the external adapter. The finite Lean core introduced below formalizes parameter indices, computed dependencies, and caching, while units and high-level observation semantics remain in the surrounding validated context.

\subsection{Point densities, bin yields, and averages}
\label{sec:observation}
For the selected external interface, let
\begin{equation}
 h(q,Q,y;\theta)=\frac{\dd\sigma}{\dd q^2\,\dd Q^2\,\dd y},
 \qquad [h]=\mathrm{pb}\,\mathrm{GeV}^{-4},
 \label{eq:native-density}
\end{equation}
where $q=q_T$. The native kernel includes its electroweak factors, Fourier normalization, flavor sum, and declared fiducial cuts. For bin $i$, define
\begin{align}
 S_i(\theta)&=\int_{q_i^-}^{q_i^+}\!\dd q
 \int_{Q_i^-}^{Q_i^+}\!\dd Q
 \int_{y_i^-}^{y_i^+}\!\dd y\;4qQ\,h(q,Q,y;\theta),
 \label{eq:yield}\\
 \mu_i(\theta)&=\frac{r_i}{\Delta q_i}S_i(\theta),
 \qquad \Delta q_i=q_i^+-q_i^- .
 \label{eq:bin-average}
\end{align}
Here $r_i$ is the source rapidity multiplicity. The factors $2q$ and $2Q$ convert the two squared-momentum measures. Mass and rapidity are integrated, while only $q$ is averaged. Therefore $S_i$ has units pb and $\mu_i$ has units pb/GeV. For this source $r_i=2$ combines the positive and negative rapidity intervals; it is not an azimuthal factor.

The controlled alternative replaces only the $q$ average by a midpoint prediction:
\begin{equation}
 \mu_i^{\rm mid}(\theta)=r_i\int_{Q_i^-}^{Q_i^+}\!\dd Q
 \int_{y_i^-}^{y_i^+}\!\dd y\;4q_i^cQ\,h(q_i^c,Q,y;\theta),
 \label{eq:midpoint}
\end{equation}
with $q_i^c=(q_i^-+q_i^+)/2$. Equations~\eqref{eq:bin-average} and \eqref{eq:midpoint} have identical output units and vector dimensions. Their equality would require an additional approximation or property of the spectrum. Accordingly, the constructor \softcode{ObservedFitProblem} rejects the midpoint operation when a finite-bin average is requested. In the failure test, this occurs before any backend or optimizer call. An explicitly unchecked test path is used only to quantify the consequence.

The observation definition must also accompany a reporting change. For an invertible linear conversion $R$ of an already defined response vector, transporting the entire statistical object,
\begin{equation}
 (\mu,d,\Sigma)\longmapsto(R\mu,Rd,R\Sigma R^{\mathsf T}),
 \label{eq:stat-transport}
\end{equation}
preserves the Gaussian quadratic objective. Multiplying predictions alone does not. This exact coordinate change is different from replacing an integral by a point value, and integration itself is generally information-losing.

\subsection{State and parameter dependence}
The native backend has process-global state. The adapter exclusively owns its process, serializes calls, and rebinds the complete nonperturbative vector at each parameter proposal. Only the selected coordinate is fitted; the other coordinates, PDF member, configuration, and precision settings remain fixed and identified. A configuration or provider change retires the session. Direct out-of-band mutations are outside the supported interface.

This state policy is relevant to formal interpretation. A callback that returns a number is not automatically an immutable constant: a native response that depends on the current nonperturbative vector must retain that dependency. Context identities and tests can detect supported changes, but a label cannot establish the purity of an arbitrary foreign function. The preservation theorem is conditional on a fixed pure interpretation, and the native adapter supplies separately audited execution evidence for its binding.

\section{Finite representations and dependency-safe plans}
\label{sec:finite}
\subsection{Signature, syntax, and interpretation}
Fix an immutable signature with atom symbols $\mathcal C$, unary symbols $\mathcal U$, and binary symbols $\mathcal B$. For $n$ parameters the source language is
\begin{equation}
\begin{split}
 e\in E_n\;::={}&\Var(i)\mid\Atom(c)\mid\Unary(u,e)\\
 &\mid\Binary(b,e_1,e_2),\qquad i\in\mathrm{Fin}(n).
\end{split}
\label{eq:grammar}
\end{equation}
An output vector is an ordered tuple of such expressions. Finite sums are explicit binary trees. The syntax preserves operation order rather than presupposing associativity of floating-point arithmetic.

A fixed pure interpretation $M$ on a set $A$ assigns
\begin{equation}
 M_{\mathcal C}:\mathcal C\to A,\quad
 M_u:A\to A,\quad M_b:A\times A\to A.
 \label{eq:interpretation}
\end{equation}
For an environment $\theta\in A^n$, define
\begin{align}
 \den{\Var(i)}_M(\theta)&=\theta_i,\nonumber\\
 \den{\Atom(c)}_M(\theta)&=M_{\mathcal C}(c),\nonumber\\
 \den{\Unary(u,e)}_M(\theta)&=M_u(\den e_M(\theta)),\nonumber\\
 \den{\Binary(b,e_1,e_2)}_M(\theta)
 &=M_b(\den{e_1}_M(\theta),\den{e_2}_M(\theta)).
 \label{eq:denotation}
\end{align}
Structural recursion gives an interpretation, and induction shows it is the unique extension satisfying these clauses. This is the usual free-term-algebra property applied to the implemented grammar. It specifies what a representation means without identifying a numerical response vector with a representable functor.

The theorem below quantifies over every fixed $M$. Addition, multiplication, and exponentiation are particular bindings; no ring or exponential law is assumed. A frozen binary64 coefficient may be treated as its exact dyadic value in a real interpretation. The distinction between that interpretation and the numerical realization of an exponential remains explicit. A native function of a fitted parameter is a primitive application to that parameter, not an atom merely because its output is a floating-point number.

\subsection{Computed dependencies and variation slices}
The support $\Dep(e)$ is computed from syntax:
\begin{align}
 \Dep(\Var(i))&=\{i\},\qquad \Dep(\Atom(c))=\varnothing,\nonumber\\
 \Dep(\Unary(u,e))&=\Dep(e),\nonumber\\
 \Dep(\Binary(b,e_1,e_2))&=\Dep(e_1)\cup\Dep(e_2).
 \label{eq:dependencies}
\end{align}
The analysis is conservative. An occurrence of $g$ remains a dependency even when a particular interpretation makes $g-g$ vanish. Rejecting such an optimization is permitted; removing a genuine dependency without proof is not.

\begin{lemma}[Dependency soundness]
\label{lem:dependency}
If $\theta$ and $\eta$ agree at every index in $\Dep(e)$, then $\den e_M(\theta)=\den e_M(\eta)$ for every fixed interpretation $M$.
\end{lemma}
\begin{proof}
Induct on $e$. A variable uses the agreement hypothesis and an atom is fixed. In a unary expression the child values agree by induction, and applying the same function preserves equality. In a binary expression, agreement on the union restricts to each child; the induction hypotheses and congruence for the fixed binary function finish the proof.
\end{proof}

A mask $V:\mathrm{Fin}(n)\to\{0,1\}$ identifies coordinates permitted to vary. For a reference environment $\theta_0$, define its variation slice
\begin{equation}
 S(V,\theta_0)=\{\theta:\theta_i=\theta_{0i}\text{ whenever }V_i=0\}.
 \label{eq:slice}
\end{equation}
The physical fitting domain may further restrict this slice. Define $\cacheable(V,e)$ by recursively checking that every variable occurrence has $V_i=0$; compound nodes use conjunction. Equivalently,
\begin{equation}
 \cacheable(V,e)\iff\Dep(e)\cap\{i:V_i=1\}=\varnothing.
 \label{eq:cacheable}
\end{equation}
Lemma~\ref{lem:dependency} immediately yields
\begin{equation}
 \cacheable(V,e)\ \Longrightarrow\
 \den e_M(\theta)=\den e_M(\theta_0),\quad\theta\in S(V,\theta_0).
 \label{eq:safe-cache}
\end{equation}
The interpretation, immutable inputs, and numerical geometry must also remain fixed. Independence from fitted coordinates does not permit reuse after a PDF, quadrature, or precision change.

\subsection{A reference-point test that misses a fitted sensitivity}
\label{sec:cache-example}
Consider the explicit fixture
\begin{align}
 U(b;g)&=\exp\!\left[R-gb^2\ln(Q/Q_A)\right],\nonumber\\
 F(a,g)&=U(b;g)\exp(-ab^2).
 \label{eq:cache-model}
\end{align}
An incorrect plan substitutes $U(b;g_\star)$ but continues to fit $g$. At $g=g_\star$, its value agrees with the correct function for every admitted kinematic point. Differentiation nevertheless gives
\begin{equation}
 \left.\partial_g F_{\rm good}-\partial_g F_{\rm frozen}\right|_{g_\star}
 =-b^2\ln(Q/Q_A)F(a,g_\star).
 \label{eq:lost-derivative}
\end{equation}
The same cancellation of the boundary derivative holds for a boundary $F_A(a,g)$ that also depends on $g$.

\begin{figure}[t]
\centering\includegraphics[width=\columnwidth]{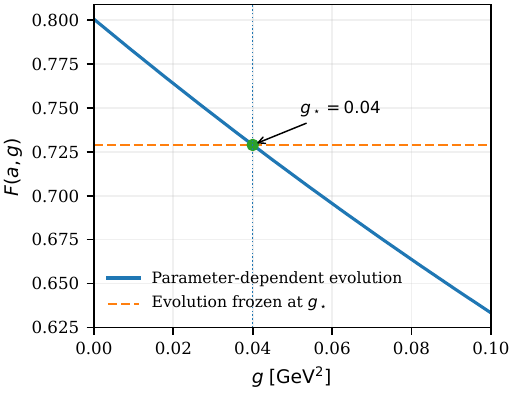}
\caption{The fitted-kernel caching counterexample of Eq.~\eqref{eq:cache-model}. Both plans agree at $g_\star$, but the frozen evolution factor removes the fitted sensitivity. Curves are direct evaluations of the stated analytic fixture, not a new extraction or an observed defect in external software.}
\label{fig:cache}
\end{figure}

With $b=1.3\GeV^{-1}$, $Q=20\GeV$, $Q_A=5\GeV$, $R=0.2$, $a=0.25\GeV^2$, and $g_\star=0.04\GeV^2$, the common value is $0.7289038135$. The correct derivative is $-1.7077031664\GeV^{-2}$, whereas the frozen derivative is zero. At $g=0.05\GeV^2$, the values are $0.7120252722$ and $0.7289038135$. Figure~\ref{fig:cache} displays this specified analytic example.

The checker finds $g\in\Dep(U)$ and rejects caching $U$ when $g$ varies, before any primitive is initialized. Declaring $g$ fixed makes a different interface on which the cache can be valid; a later attempt to vary $g$ then violates the fixed-coordinate condition. A conventional derivative or off-reference test can detect this injected error. The formal advantage is the all-environment guarantee for every accepted transformation in the supported language, rather than superiority to every possible unit-test suite.

\subsection{Plans and the concrete checker}
The plan language has live and cached source subexpressions plus lifted primitive operations:
\begin{equation}
 p::=\Live(e)\mid\Cache(e)\mid U_p(u,p)\mid B_p(b,p_1,p_2).
 \label{eq:plan}
\end{equation}
A cache retains the expression that produces its value. It does not accept an arbitrary numerical table asserted to equal that expression. Erasure removes the live/cache annotations and reconstructs the source tree. The safety predicate accepts a live term, accepts a cache precisely when Eq.~\eqref{eq:cacheable} holds, and recursively checks compound plans. The implemented Boolean check is
\begin{equation}
 \PlanCheck(V,e,p)=[\erase(p)=e]\wedge\safe(V,p).
 \label{eq:checker}
\end{equation}
The equality in the first conjunct is decidable syntactic equality, not a floating-point comparison or an untrusted semantic oracle.

Plan evaluation uses both reference and current environments:
\begin{align}
 \Eval_M(\Live(e);\theta_0,\theta)&=\den e_M(\theta),\nonumber\\
 \Eval_M(\Cache(e);\theta_0,\theta)&=\den e_M(\theta_0),
 \label{eq:plan-evaluation}
\end{align}
with the same primitive functions at compound nodes.

\begin{theorem}[Preservation by the concrete validator]
\label{thm:check}
For every fixed pure interpretation $M$, source $e$, plan $p$, mask $V$, reference $\theta_0$, and $\theta\in S(V,\theta_0)$,
\begin{equation}
 \PlanCheck(V,e,p)=\mathrm{true}
 \quad\Longrightarrow\quad
 \Eval_M(p;\theta_0,\theta)=\den e_M(\theta).
 \label{eq:preservation}
\end{equation}
\end{theorem}
\begin{proof}
Induct on the safe plan. A live term is immediate. A cached term follows from Eq.~\eqref{eq:safe-cache}. Unary and binary cases follow by induction and application of the same fixed functions. Thus a safe plan evaluates to its erasure. The first conjunct of Eq.~\eqref{eq:checker} identifies that erasure with $e$.
\end{proof}

This is the theorem implemented as \softcode{check\_sound}. Its conclusion is derived from the concrete checker; it is not a premise carried by the candidate plan. The proposer can therefore be an ordinary program, an optimizer, or an AI system without being trusted to establish preservation. The checker is intentionally incomplete: it does not authorize reassociation, exponential splitting, finite-sum permutations, changed quadrature, or substitutions between physical prescriptions. Each such extension would require a separately specified rule and proof.

\section{Categorical composition and Lean verification}
\label{sec:category}
\subsection{Interfaces and admitted substitutions}
A local preservation theorem is useful in a larger calculation only when its assumptions survive composition. Fix the signature and primitive context of Sec.~\ref{sec:finite}. An object is a parameter interface
\begin{equation}
 X=(n,V_X).
 \label{eq:interface}
\end{equation}
For $Y=(m,V_Y)$, an arrow $f:X\to Y$ is an ordered tuple $(f_0,\ldots,f_{m-1})$ with $f_j\in E_n$, satisfying the computed restriction
\begin{equation}
 V_Y(j)=0\quad\Longrightarrow\quad\cacheable(V_X,f_j).
 \label{eq:admissibility}
\end{equation}
A target coordinate declared fixed can therefore depend only on source-fixed coordinates. The implementation's proof field establishes these Boolean checks; it does not postulate equality of complete responses.

For a substitution $\sigma:i\mapsto\sigma_i$, define $e[\sigma]$ by replacing variable leaves and retaining the other constructors. Structural induction gives
\begin{align}
 e[\Var]&=e,\nonumber\\
 (e[\sigma])[\tau]&=e[i\mapsto\sigma_i[\tau]],\nonumber\\
 \den{e[\sigma]}_M(\theta)
 &=\den e_M\bigl(i\mapsto\den{\sigma_i}_M(\theta)\bigr).
 \label{eq:subst-laws}
\end{align}
Identity is the variable tuple, and composition is
\begin{equation}
 (g\circ f)_k=g_k[f_0,\ldots,f_{m-1}].
 \label{eq:arrow-composition}
\end{equation}
Admissibility is closed under this operation: a cacheable target expression uses only fixed intermediate coordinates, and their admitted substitutions use only fixed source coordinates. Induction on the expression proves that its substituted expression is cacheable. Equation~\eqref{eq:subst-laws} then establishes both identity laws and associativity. This constructs a concrete variation-aware term category.

For each fixed interpretation $M$, define
\begin{equation}
 I_M(X)=A^n,\qquad
 I_M(f)(\theta)_j=\den{f_j}_M(\theta).
 \label{eq:functor}
\end{equation}
The interpretation is a functor into functions between parameter environments:
\begin{equation}
 I_M(\mathrm{id}_X)=\mathrm{id}_{A^n},\qquad
 I_M(g\circ f)=I_M(g)\circ I_M(f).
 \label{eq:functor-laws}
\end{equation}
These are standard categorical laws~\cite{Riehl2016}; the substantive choice here is the interface information, concrete syntax, and admissibility rule they organize. Changing a provider or primitive interpretation is not automatically an arrow in this category. Such a change needs its own comparison construction.

\subsection{Transported reference environments}
\begin{lemma}[Preservation of the variation slice]
\label{lem:slice}
For an admitted $f:X\to Y$,
\begin{equation}
 I_M(f)\bigl(S(V_X,\theta_0)\bigr)
 \subseteq S\bigl(V_Y,I_M(f)(\theta_0)\bigr).
 \label{eq:slice-transport}
\end{equation}
\end{lemma}
\begin{proof}
If target coordinate $j$ is fixed, Eq.~\eqref{eq:admissibility} makes $f_j$ cacheable on the source slice. Equation~\eqref{eq:safe-cache} equates its values at $\theta$ and $\theta_0$. This holds for every target-fixed coordinate.
\end{proof}

Let $p_f$ and $p_g$ be componentwise accepted plans for arrows $f:X\to Y$ and $g:Y\to Z$. Writing $\Eval$ componentwise, the next theorem identifies the correct initialization of the second plan.

\begin{theorem}[Composition of checked stages]
\label{thm:composition}
Set $\eta_0=I_M(f)(\theta_0)$. For every $\theta\in S(V_X,\theta_0)$,
\begin{equation}
\begin{split}
 \Eval_M\bigl(p_g;\eta_0,\Eval_M(p_f;\theta_0,\theta)\bigr)
 =I_M(g\circ f)(\theta).
\end{split}
\label{eq:checked-composition}
\end{equation}
\end{theorem}
\begin{proof}
Theorem~\ref{thm:check} identifies the first plan's output with $I_M(f)(\theta)$. Lemma~\ref{lem:slice} places that output in the target slice referenced at $\eta_0$, so Theorem~\ref{thm:check} applies to the second plan. The interpretation law~\eqref{eq:functor-laws} gives the composite.
\end{proof}

The reference condition matters even when both plans separately pass their checks. In the implemented example,
\begin{equation}
 f(x,y)=(x+2,y^2),\qquad g(u,v)=uv,
 \label{eq:composition-example}
\end{equation}
$x$ varies and $y$ is fixed. The intermediate interface has $u$ varying and $v$ fixed. Initializing the second stage with $v_0=y_0$ instead of $v_0=y_0^2$ generally gives the wrong product, although its dimensions and parameter names can look compatible. The correct initialization is $f(x_0,y_0)$; the runtime rejects the untransported reference with \softcode{fixed\_parameter\_changed}. At $y_0=2$ and $x=1$, for example, the correct composite is $12$, whereas the unrelated fixed coordinate would give $6$. These are illustrative evaluations of Eq.~\eqref{eq:composition-example}, not another inference experiment.

The category thereby determines when local guarantees can be reused in a larger pipeline. A reporting conversion may be invertible, while an observation such as $(x_1,x_2)\mapsto x_1+x_2$ is not. Numerical plans with different caching choices retain their own syntax and witnesses: the present theorem is semantic composition of checked stages, not an unimplemented strict category of arbitrary optimized plans.

\subsection{Concrete formalization and proof scope}
The formal library implements \softcode{Expr n}, \softcode{Plan n}, a fixed \softcode{Model}, \softcode{Interface}, and admitted \softcode{Arrow} records. The category and functor laws are proved directly, without a mathlib dependency. Lean 4.21.0 checks dependency soundness, cache constancy, the concrete validator, substitution and interpretation, fixed-slice transport, and Theorem~\ref{thm:composition}. Twenty-four library declarations, including helpers and explicit example results, are listed in Appendix~\ref{app:inventory}.

The audited dependencies are the standard Lean axioms \softcode{propext} and \softcode{Quot.sound}; some declarations use neither. The result should therefore not be called axiom-free. The evidence includes theorem statements and \softcode{\#print axioms} output, not merely searches for prohibited source tokens. No \softcode{sorryAx}, added project axiom, or native-evaluation shortcut closes the proof obligations. These distinctions follow the separation between a valid proof, the meaning of its statement, and its logical dependencies~\cite{LeanValidation,LeanAxioms}.

\subsection{From serialized plans to accepted execution}
\label{sec:certificate}
The exchange representation contains topologically ordered expression and plan tables, explicit parameter and primitive identifiers, root indices, and canonical finite numerical inputs. The decoder rejects unsupported fields, duplicate keys, invalid indices, cycles or forward references, noncanonical/nonfinite numerical values, and unknown primitives. These decoder properties are tested rather than formally proved.

The function \softcode{prepare\_certified} seals a canonical copy of the submitted representation, validates its structure and primitive registry, emits literal Lean definitions, and requests a fresh certificate. Each positive certificate contains ordinary kernel-reduced acceptance followed by the soundness theorem. Schematically, the literal proof has the form
\begin{lstlisting}
theorem accepted :
  check varies source candidate = true := by decide
\end{lstlisting}
The actual generated certificate then instantiates \softcode{check\_sound} for those definitions. A negative certificate proves that the frozen-kernel candidate returns false. After success, the same sealed representation is decoded for evaluation. Cache values are computed from retained source subexpressions at the reference state; a caller cannot inject an unrelated cache table under the supported interface.

Table~\ref{tab:certificates} gives the complete selected roots checked in the reviewed execution. Both the supplied and locally regenerated snapshots received certificates, so historical repeated checks are not counted as extra physical cases. All remaining output roots have runtime bridge tests; the universal library theorem does not imply that an unexamined root has itself received a literal acceptance certificate.

\begin{table*}[tbp]
\caption{Selected complete literal plans receiving acceptance certificates. Counts describe the recorded expression/plan tables. One complete output root per listed case was checked in each of two snapshot sets. The external-geometry case uses an explicitly analytic point-kernel fixture; its certificate does not verify native arTeMiDe.}
\label{tab:certificates}
\begin{tabularx}{\textwidth}{@{}>{\raggedright\arraybackslash}p{.25\textwidth}rrY@{}}
\toprule
Case & Source nodes & Plan nodes & Interpreted and tested scope\\
\midrule
Fitted-kernel example & 11 & 11 & Correct cache plan, explicit unsafe alternative, reference and off-reference evaluations.\\
Diagnostic TMD, chart A & 4,996 & 4,228 & Full first response built from the actual finite-window coefficient arrays.\\
Diagnostic TMD, chart B & 6,148 & 5,380 & Full first response from the separately assembled transported-boundary route.\\
External observation geometry & 7,917 & 7,170 & Full first bin with actual nodes, Jacobians, averaging, and a parameter-explicit analytic kernel.\\
\bottomrule
\end{tabularx}
\end{table*}

The initial large certificate exhausted the default Lean thread stack. The identical certificate passed with the per-invocation setting \softcode{--tstack=65536}; its source bytes and proof method were unchanged. A dense integration regression retains that workload. This is a resource adjustment, not a relaxation of proof checking.

An independent literal/table audit compared serialized rows with emitted Lean terms, root identifiers, masks, source identities, certificate digests, and recorded evaluations. These checks establish correspondence for the delivered artifacts. They are not a general verified decoder, Python interpreter, or foreign-function interface. The mathematical interpretation remains fixed and pure: an opaque callback reading an undeclared mutable fitted parameter violates the premise rather than receiving protection from its name or hash.

\section{External Drell--Yan application}
\label{sec:external}
\subsection{Pinned physics and the one-parameter study}
The external calculation uses arTeMiDe v2.05 and its bundled SV19 model. The Drell--Yan configuration is \path{DY_nnlo/const-NNPDF31_NNLO}~\cite{ScimemiVladimirov2020,Artemide205}. The pinned engine and DataProcessor commits are recorded in Appendix~\ref{app:reproduction}. NNPDF3.1 NNLO member 0 and its native LHAPDF coupling supply the collinear input~\cite{NNPDF31,LHAPDF6}. The native momentum-weighted PDF convention stays inside the external kernel; it is not silently substituted for the diagnostic model's interface returning $f_i(x,\mu)$.

The selected process is proton--proton neutral-current Drell--Yan, with photon, $Z$, and interference contributions and the native signed quark/antiquark sum through bottom. The source's coefficient inventory uses the cusp anomalous dimension through order three, noncusp and rapidity coefficients through order two, and hard and matching functions through $\alpha_s^2$. The resummed rapidity and $\zeta$-line settings are order two, and evolution solution type 3 uses the native $\zeta$ prescription. The exact-$x$ option, electroweak parameters, thresholds, and disabled $\pi^2$ resummation are retained. This inventory, rather than an inferred logarithmic label, defines the calculation. The low-$q_T$ prediction has no matched large-$q_T$ $Y$ term.

In the pinned SV19 evolution model,
\begin{equation}
 b_*^{\rm SV}=\frac{b}{\sqrt{1+b^2/B^2}},\qquad
 \mathcal D(\mu,b)=\mathcal D_{\rm resum}(\mu,b_*^{\rm SV})+c_0bb_*^{\rm SV}.
 \label{eq:sv19}
\end{equation}
The complete nine-component nonperturbative vector is rebound for every proposal. Only $c_0$ varies. Its generating value and local search interval are
\begin{equation}
 c_{0\star}=0.029551\GeV^2,\qquad
 0.001\le c_0/\mathrm{GeV}^2\le0.10.
 \label{eq:c0domain}
\end{equation}
The other components are fixed to the published-mean row of the pinned model input, not its initialization row. This is a restricted synthetic inference experiment, not the published global fit. In particular, $c_0$ is not identified with the parameter $g_2$ of the separate diagnostic model.

\subsection{Geometry, cuts, and numerical routes}
The geometry is taken from the source dataset corresponding to ATLAS $8$-TeV dilepton spectra~\cite{ATLAS2016,ArtemideData}. It has $66<Q<116\GeV$, $0<y<0.4$ with multiplicity two, and native cuts $p_T^\ell>20\GeV$, $|\eta_\ell|<2.4$. The source-ordered selection $q_T^+/Q^-\le0.25$ gives bins
\begin{equation}
 [0,2],\ [2,4],\ [4,6],\ [6,8],\ [8,10],\ [10,13],\ [13,16]\GeV.
 \label{eq:bins}
\end{equation}
The first five bins train the local fit; the final two are held out. Selection, numerical budgets, and inference settings were frozen before inspecting the mismatch's fit effect.

Three routes have distinct validation roles. N1 calls the native bin interface and applies source theory factors. N2 sends the same request through the typed adapter, including off-central and change/reset probes. Their equality validates the adapter and state binding. N3 independently integrates the native point response using Eq.~\eqref{eq:bin-average}. It is independent observation quadrature sharing the native physics kernel, not an independent QCD engine.

The N3 tensor Gauss--Legendre rule uses orders $(8,8,4)$ in $(q,Q,y)$ and mass subdivisions at $80$, $86$, $89$, $91$, $93$, $96$, and $102\GeV$ to resolve the resonance region. Independent refinements double one order at a time. The nominal native settings use a minimum of 32 transverse Simpson sections and $2\times10^{-7}$ mass/rapidity tolerance; the wider held-out bins receive 34 sections from the native rule. A separate run uses minimum 64 and $10^{-7}$ tolerance. The Fourier tolerance is $10^{-6}$ in both. An earlier pilot with fewer sections failed a numerical comparison, and its retained failure motivated increased resolution, not relaxed acceptance.

Generating means come from N1. Fits use the faster N3 observation integral, checked against N1 at the generating and selected fitted parameters. The midpoint alternative keeps the mass and rapidity integrations and all source factors, changing only the $q_T$ observation operation. Table~\ref{tab:routes} prevents the word ``independent'' from obscuring what each comparison establishes.

\begin{table}[t]
\caption{Roles of the external comparison routes.}
\label{tab:routes}
\small
\begin{tabularx}{\columnwidth}{@{}p{.12\columnwidth}Y@{}}
\toprule
Route & Calculation and evidence\\
\midrule
N1 & Direct native integrated-bin interface; synthetic generating means.\\
N2 & The same native request through validated types; adapter and state checks.\\
N3 & Separate integration of the native point kernel; independent observation quadrature used for fitting.\\
Fault & N3 mass/rapidity integration with a $q_T$ midpoint substituted for the finite-bin average; isolated unchecked test only.\\
\bottomrule
\end{tabularx}
\end{table}

\subsection{Frozen inference design}
Let $t_i$ be the native generating means and $T=\max_{i\in\mathrm{train}}|t_i|$. Define
\begin{equation}
 \sigma_i=\max(0.02|t_i|,10^{-4}T),\qquad
 \Sigma=\operatorname{diag}(\sigma_i^2).
 \label{eq:sigma}
\end{equation}
The covariance is fixed across routes. A single saved PCG64 realization gives $d_i=t_i+\sigma_i z_i$; the seed is 2026102006 and the realized numbers are archived. The noiseless and fixed-noise data are each fitted with the correct and midpoint maps. The final two bins are compared to their known means rather than used in the objective. No experimental measured cross sections or covariance are fitted.

The solver is SciPy's bounded trust-region reflective least-squares routine~\cite{SciPy}, with linear loss, a centered finite-difference Jacobian, and dimensionless coordinate $u=(c_0-0.001)/0.099$. Five starts have $u=0.125$, $0.25$, $0.50$, $0.75$, and $0.875$. The three stopping tolerances are $10^{-7}$, with at most 50 function evaluations. The derivative uses a chart step $10^{-3}$; a separate local-sensitivity comparison checks step variation. All twenty starts terminate successfully, and selected solutions are interior. This multistart evidence does not constitute a theorem of global minimization.

The weighted training sensitivity at truth has rank one and norm $61.1386\GeV^{-2}$, giving the local scale
\begin{equation}
 s_{c_0}=\left[\sum_{i\in\mathrm{train}}
      \sigma_i^{-2}\bigl(\partial_{c_0}\mu_i\bigr)^2\right]^{-1/2}
 =0.0163563\GeV^2.
 \label{eq:sensitivity}
\end{equation}
This scale helps display the effect; it is not a calibrated significance assigned to a model-error displacement.

\subsection{Detection and measured repair}
\begin{figure}[t]
\centering\includegraphics[width=\columnwidth]{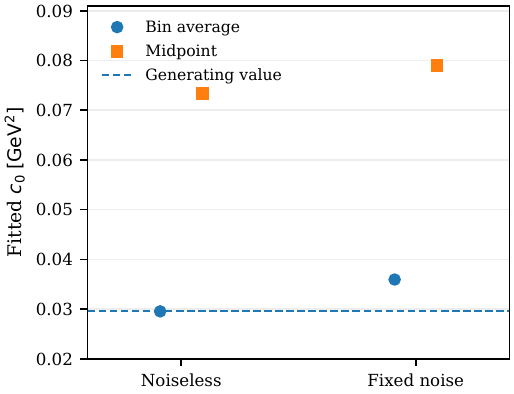}
\caption{The fitted SV19 evolution parameter for the two frozen synthetic data vectors and two observation maps. The dashed line marks the generating value. A midpoint construction can shift the parameter substantially without an obviously catastrophic residual. The symbols are selected fits, not an ensemble or confidence region.}
\label{fig:external-parameters}
\end{figure}

The production fit constructor rejects the point observation before any backend or optimizer invocation. The isolated unchecked path then supplies the measured counterexample in Table~\ref{tab:external-fits} and Fig.~\ref{fig:external-parameters}. For noiseless data the corrected estimate differs from truth by about $1.62\times10^{-5}\GeV^2$ within the achieved numerical resolution. The midpoint estimate is displaced by $0.043864\GeV^2$ relative to the corrected fit, or $2.68s_{c_0}$, while retaining a minimum $\chi^2$ of $1.56$.

\begin{table*}[tbp]
\caption{External synthetic fits. All $c_0$ values are in $\mathrm{GeV}^2$; the generating value is $0.029551$. The midpoint route deliberately bypasses the rejecting observation constructor. No statistical error bars are inferred from the two displayed realizations.}
\label{tab:external-fits}
\begin{tabular}{@{}lrrrr@{}}
\toprule
Data & Correct bin average $c_0$ & Midpoint $c_0$ & Correct $\chi^2_{\min}$ & Midpoint $\chi^2_{\min}$\\
\midrule
Noiseless & 0.029535 & 0.073399 & $8.95\times10^{-7}$ & 1.56283\\
Fixed noise & 0.035936 & 0.079022 & 2.38283 & 5.22883\\
\bottomrule
\end{tabular}
\end{table*}

\begin{figure}[t]
\centering\includegraphics[width=\columnwidth]{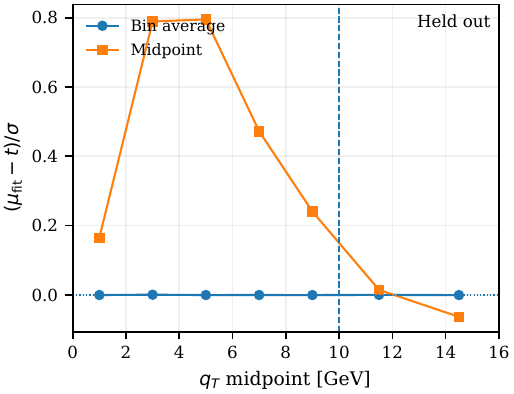}
\caption{Noiseless external response residuals after fitting the first five bins. Each value is divided by the frozen synthetic standard deviation. The last two bins, to the right of the vertical boundary, are held out. Lines connect saved bin values and do not represent a new continuous-spectrum calculation.}
\label{fig:external-residuals}
\end{figure}

The fixed-noise parameter displacement is $0.043086\GeV^2$, with a $\chi^2$ increase of $2.8460$. The two held-out bins discriminate weakly: in the noiseless comparison their mean residual norms are approximately $0.00036$ for the correct map and $0.0647$ for the midpoint map in the declared comparison units. Figure~\ref{fig:external-residuals} shows why the parameter displacement, not a claim of strong held-out separation, is the main result.

The numerical acceptance combines a componentwise budget with relative part $10^{-4}$ and absolute part $10^{-8}$ times the native batch maximum, and a whitened norm $0.05$. Individual refinement differences have budget $0.05/3$. Across the selected comparisons, the largest component-budget ratio is $0.86094$, the largest whitened integration change is $0.0044023$, and the largest native-refinement change is $0.0012731$. The observed objective increases exceed the respective empirical two-route objective-perturbation budgets $0.13011$ and $0.38803$; Appendix~\ref{app:error} states the inequality used to calculate them.

The experiment therefore resolves the repair within the specified numerical checks. It does not supply a rigorous integration bound. The fault is a controlled analysis-interface misuse of a real public calculation, not a discovered defect in arTeMiDe, ATLAS, or the published SV19 extraction. A typed contract makes the mismatch explicit and prevents that supported misuse before fitting; its measured consequence explains why the distinction is worth retaining.

\section{Numerical, inference, and computational validation}
\label{sec:validation}
\subsection{A transparent two-parameter regression}
A separate diagnostic supplies a tractable setting for independent scalar assembly, boundary transport, derivative checks, and repeated inference. Its local inventory is tree-level hard and small-$b$ matching coefficients with one-loop evolution generators and a completed integrable kernel, using CT18NNLO member 0 and its coupling~\cite{CT18,LHAPDF6}. The external PDF's metadata is retained as input provenance; it does not assign higher orders to coefficients implemented locally. This diagnostic must not be confused with the external SV19 settings of Sec.~\ref{sec:external}.

The finite-window target is
\begin{equation}
 W(Q,q_T;\theta)=\frac{1}{2\pi}\int_{10^{-4}}^{8}\!b\,\dd b\,
 J_0(bq_T)\,\mathcal L(x_a,x_b,b;Q,\theta),
 \label{eq:diagnostic-W}
\end{equation}
where both endpoints have units $\mathrm{GeV}^{-1}$ and $\mathcal L$ is the two-beam $u,d,s$ luminosity including both quark/antiquark assignments and charge weights. The response is per $\dd^2q_T$, in $\mathrm{GeV}^{-2}$, without a cross-section prefactor or $Y$ term. The integration window is part of its definition. Earlier endpoint shifts were below the numerical resolution and do not supply a quantitative omitted-tail bound.

The parameters are $\theta=(a,g_2)$, with $\theta_\star=(0.25,0.04)\GeV^2$ and domain $[0.10,0.50]\times[0,0.10]\GeV^2$. At $\sqrt{s}=200\GeV$ and $y=0.3$, six training responses use $Q=20,40\GeV$ and $q_T/Q=0.025,0.05,0.10$. Three mean predictions at $Q=30\GeV$ are held out. Their momentum fractions change with $Q$, so this is a prediction at new kinematics rather than an isolated fixed-$x$ evolution test.

The two reference presentations are connected by
\begin{equation}
 \Psi(a,g_2)=(\beta,g_2)=(a+g_2\ln2,g_2).
 \label{eq:chart}
\end{equation}
The entire boundary family, its fitted kernel dependence, and its domain are transported. In particular, the second domain is the exact image parallelogram, not an unrelated rectangle. Appendix~\ref{app:diagnostic} states the complete formulas. With the same observations and covariance, the forward identity implies
\begin{equation}
 \chi_A^2=\chi_B^2\circ\Psi,\qquad
 \Psi\!\left(\operatorname*{argmin}_{\Theta_A}\chi_A^2\right)
 =\operatorname*{argmin}_{\Theta_B}\chi_B^2.
 \label{eq:argmin}
\end{equation}
This standard change-of-presentation result does not assume a unique optimum. It is a positive regression of complete transport, rather than the principal novelty theorem of this work.

Separate production and scalar-reference assemblies use distinct outer quadratures and independent accumulation of the scale moments. The same primitive physics inputs are documented as shared. Checks cover parameter-domain corners, selected minima, held-out responses, derivatives, and both transport paths. All twelve required A/B fit pairs satisfy the recorded prediction, parameter, and objective budgets. The noiseless approved fit has $\chi^2_{\min}=8.67\times10^{-13}$. An optional prescribed noncusp-coefficient control retains integrable transport but gives $\chi^2_{\min}=76.60846$ after a noiseless refit, distinguishing evolution consistency from equality of models. It is a stress test, not an uncertainty band or a detected error in a published calculation.

\subsection{Closure and the paired coverage diagnosis}
\label{sec:coverage}
The diagnostic covariance uses Eq.~\eqref{eq:sigma} with its own generating responses. All realized noise arrays are saved. Two hundred replicas are fitted in the canonical chart, with ten fixed replicas also fitted through the transported chart. Five starts are retained for each fit. There are no failed selected fits; two canonical replicas reach $g_2=0$. Ordinary interior inverse-Fisher covariance bands are suppressed at parameter boundaries rather than reported as zero uncertainty.

The fixed nominal joint likelihood-ratio thresholds are $2.278868566$ and $5.991464547$ for probabilities 0.68 and 0.95 with two degrees of freedom. These thresholds define the procedure being measured; they do not by themselves establish finite-sample coverage for a bounded nonlinear model~\cite{Cowan2011}. The observed inclusion fractions are $137/200=68.5\%$ and $184/200=92.0\%$. Their 95\% Wilson intervals are $[61.765\%,74.538\%]$ and $[87.401\%,95.016\%]$~\cite{Wilson1927}. Nominal 95\% lies only just inside the second interval.

The empirical 95\% quantile of the nonlinear likelihood ratio is $6.82395$. To diagnose this result without changing the data, let
\begin{equation}
 G=\Sigma^{-1/2}J_\star D,\qquad
 D=\operatorname{diag}(0.40,0.10)\GeV^2,
 \label{eq:whitened-jacobian}
\end{equation}
where $J_\star$ is the derivative at truth. Its singular values are $87.7563$ and $5.13503$, with condition number $17.0897$ and canonical parameter correlation $-0.990937$. Both local directions are resolved; the correlation describes their coordinate geometry.

\begin{figure}[t]
\centering\includegraphics[width=\columnwidth]{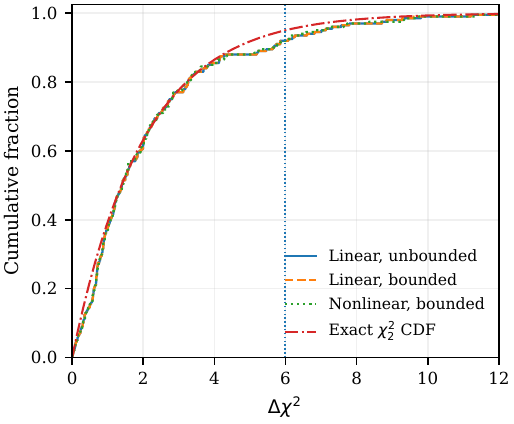}
\caption{Empirical distributions of the paired likelihood-ratio statistics and the exact $\chi^2_2$ cumulative distribution for the unbounded linear control. The vertical line is the fixed nominal 95\% threshold. The tail-heavy saved realization is already present in the exact linear statistic; this comparison diagnoses the sample, not universal nonlinear coverage.}
\label{fig:coverage}
\end{figure}

For a thin QR decomposition $G=QR$, the unbounded linear least-squares improvement for standard-normal noise $z$ is
\begin{equation}
 \Delta\chi^2_{\rm lin}=\|Q^{\mathsf T}z\|_2^2\sim\chi_2^2.
 \label{eq:linear-stat}
\end{equation}
This distribution is exact for fixed rank-two $G$, irrespective of a nonsingular parameter-coordinate change. Correlation alone therefore does not explain a shifted threshold. Applying Eq.~\eqref{eq:linear-stat} to the identical saved noise, and then adding the original parameter bounds in a paired linear calculation, gives Table~\ref{tab:coverage} and Fig.~\ref{fig:coverage}.

\begin{table}[t]
\caption{Post-hoc coverage diagnosis using the same 200 saved noise vectors. All three rows include 137 truths at the nominal 68\% threshold and 184 at the nominal 95\% threshold. The original thresholds were not recalibrated.}
\label{tab:coverage}
\small
\begin{tabularx}{\columnwidth}{@{}Yrr@{}}
\toprule
Calculation & Empirical $q_{0.95}$ & 95\% inclusion\\
\midrule
Linear, unbounded & 7.01529 & 184/200\\
Linear, original bounds & 6.97097 & 184/200\\
Nonlinear, original bounds & 6.82395 & 184/200\\
\bottomrule
\end{tabularx}
\end{table}

The unbounded linear and nonlinear calculations exclude exactly the same sixteen replicas at nominal 95\%. The same finite sample thus suffices to produce the elevated quantile; the bounds and nonlinear calculation lower its observed value. This does not establish exact coverage for the nonlinear procedure. For context, the probability of at least sixteen exclusions in 200 trials with exact inclusion probability 0.95 is about 0.0444. That is a post-hoc descriptive calculation, not a prespecified calibration or model-selection test. The retained coverage evidence remains limited to the tested truth, covariance, and candidate family.

\subsection{Bridges from the numerical calculations}
The diagnostic bridge exports the actual compiled coefficient arrays into finite source and plan tables. Frozen luminosity, PDF, scale-moment, Bessel, and quadrature factors become atoms; $a$ or $\beta$ and $g_2$ remain variable leaves. Both presentations are exported from their separately assembled factors. At four parameter points across nine responses and two charts, 72 original-versus-plan comparisons pass, with largest absolute difference $2.22\times10^{-16}$. The historical NumPy reduction and the explicit expression tree are compared numerically, not assumed bitwise identical.

For the external geometry, the finite observation takes the form
\begin{equation}
 A_i(c_0)=\frac{r_i}{\Delta q_i}
 \sum_j w_j^q w_j^Q w_j^y\,4q_jQ_j\,h_j(c_0).
 \label{eq:finite-observation}
\end{equation}
Every $h_j$ is a parameter-explicit unary primitive; other fixed inputs belong to its context. Geometry can be cached, while $h_j(c_0)$ cannot be treated as parameter independent. The bridge uses the actual seven-bin geometry and its 14,336 sample points, with an explicit analytic point-kernel fixture. Twenty-eight comparisons at four parameter values agree with the existing observation quadrature to a largest scaled difference of about $2.3\times10^{-16}$ in the returned Mac execution. The selected full first-bin certificate in Table~\ref{tab:certificates} concerns this binding, not native arTeMiDe.

The diagnostic extraction was also reproduced across Mac and Linux environments using identical saved inputs. Principal parameter differences are below $5.2\times10^{-12}\GeV^2$, all coverage memberships agree, and the largest selected objective difference is about $8.2\times10^{-14}$. Some nearly tied start identifiers differ under floating-point ordering while the scientific outputs remain within their budgets. A separate saved-array audit recomputes objectives, memberships, covariance transformations, and response projections without importing the TMD implementation. Such arithmetic consistency complements, rather than replaces, independent assembly and source-level checks.

\subsection{Same-backend checking overhead}
\label{sec:performance}
The timing comparison uses the same native point kernel, quadrature, full parameter rebinding, and cache regime on one machine. Direct and typed paths perform equal backend-call counts. Thirty paired blocks alternate which path is timed first, following two warmups per path. The recorded clock is \softcode{perf\_counter\_ns}; the native run is serial. Construction/validation, one-bin evaluation, seven-bin evaluation, and a three-proposal objective workload are measured separately. The final workload retains the recorded complete seven-bin response evaluation at each proposal, so its cost is not misrepresented as a minimal five-bin objective implementation.

\begin{table*}[tbp]
\caption{Same-backend runtime comparison in the recorded Mac environment. Times are medians per workload, and the ratio is the median of paired block ratios rather than necessarily the ratio of the two marginal medians. Native startup and Lean certificate checking are outside these timings.}
\label{tab:timing}
\begin{tabularx}{\textwidth}{@{}Yrrr@{}}
\toprule
Workload & Direct median & Typed median & Paired ratio\\
\midrule
Construction and validation & $59.44\,\mu\mathrm{s}$ & $79.81\,\mu\mathrm{s}$ & 1.3422\\
One bin, warm & $0.20336\,\mathrm{s}$ & $0.20304\,\mathrm{s}$ & 0.9976\\
Seven bins, warm & $1.54727\,\mathrm{s}$ & $1.54251\,\mathrm{s}$ & 0.9967\\
Three changing-parameter proposals & $4.65228\,\mathrm{s}$ & $4.64119\,\mathrm{s}$ & 0.9970\\
\bottomrule
\end{tabularx}
\end{table*}

\begin{figure}[t]
\centering\includegraphics[width=\columnwidth]{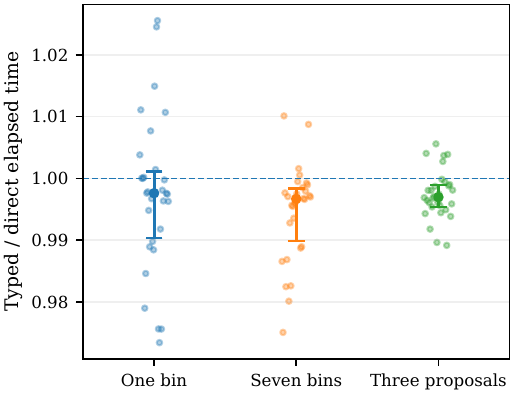}
\caption{Paired timing ratios for the numerical workloads. Small points show all 30 alternating blocks; the larger marker and bars show the median and interquartile range. Near-unity ratios indicate that numerical work dominates the observed runtime. Values below one are not interpreted as a demonstrated speedup.}
\label{fig:overhead}
\end{figure}

The median paired construction increment is $20.34\,\mu\mathrm{s}$. Numerical workload ratios are near unity (Table~\ref{tab:timing}, Fig.~\ref{fig:overhead}); their small deviations below one are not a speedup claim. These measurements concern the runtime observation layer, not formal proof generation. A literal certificate is a separate setup operation whose cost depends on term size and reduction; this work reports its workload and successful execution rather than assigning the runtime adapter's cost to Lean. Peak memory was not measured.

\section{Software architecture and reproducibility}
\label{sec:software}
\subsection{A source, a candidate plan, and an evidence record}
The implementation separates declaration, construction, proposed transformation, and evidence. Immutable records retain physical and numerical context. Expression trees retain dependency information and the order of primitive operations. A numerical compiler can propose reusable factors, but the accepted plan preserves the source and cache side conditions. Observation constructors and parameter-state guards protect the interfaces surrounding that finite representation.

These roles support both inspection and replay. Source and candidate identifiers, wire and certificate digests, primitive/input identities, selected roots, actual toolchain, and evaluation results are recorded together. A digest identifies the associated bytes; the surrounding proof or numerical comparison states what has been established about them. The source expression is produced by an identified lowering rule and then bound to the plan. Otherwise a proposer could submit an incorrect source and a faithfully preserving plan without answering the intended scientific question.

Table~\ref{tab:trust} summarizes the guarantees. The distinction between runtime and formal verification is part of the API: the runtime-only examples can operate without Lean, but a requested certificate has no silent runtime fallback. A missing toolchain or input is an explicit failure or blocked obligation.

\begin{table*}[tbp]
\caption{Evidence and assumptions. ``Exact'' refers to the stated mathematical interpretation; it does not promote numerical convergence diagnostics into certified error bounds.}
\label{tab:trust}
\begin{tabularx}{\textwidth}{@{}>{\raggedright\arraybackslash}p{.22\textwidth}>{\raggedright\arraybackslash}p{.34\textwidth}Y@{}}
\toprule
Layer & Established evidence & Remaining premise or limitation\\
\midrule
Physical source and observation & Source audit, checked runtime assembly, native/typed agreement, independent observation integration & The selected factorization/model and backend interpretation; no kernel proof of QCD.\\
Finite representation & Lean-checked dependency lemma, cache soundness, source-category and interpretation laws & Fixed pure primitive interpretation and admitted fixed-coordinate environment.\\
Selected plan instances & Literal source/plan acceptance and soundness instances for the roots in Table~\ref{tab:certificates} & Other roots are separately runtime-tested, not individually certified by implication.\\
Encoding and execution & Strict decoder, sealed-plan binding, cross-language/table audit, reference/off-reference tests & No general formal proof of JSON decoding, Python, or the foreign-function interface.\\
Numerical evaluation & Independent assembly, refined quadrature, derivative checks, native precision checks & Empirical accuracy on the stated domain; no full-transform-tail or IEEE theorem.\\
Inference & Frozen observations/covariance, controlled fits, local sensitivity, paired coverage study & Conditional on the specified model, inputs, and statistical design.\\
\bottomrule
\end{tabularx}
\end{table*}

\subsection{Reproduction tiers and release identity}
The implementation is cited by its exact version, typed-TMD v0.4.0, DOI \href{https://doi.org/10.5281/zenodo.23546800}{10.5281/zenodo.23546800}~\cite{TypedTMD2026}. The project-wide DOI \href{https://doi.org/10.5281/zenodo.23546799}{10.5281/zenodo.23546799} identifies the evolving software family; the version DOI identifies the release associated with this article. Original code is MIT-licensed. Third-party code and grids are obtained from their identified sources under their own terms rather than relicensed as part of typed-TMD.

Reproduction has four separate scopes. The lightweight runtime path exercises dependency checks, safe and unsafe caching, staged reference transport, analytic closure, and saved-array coverage. Formal replay builds the pinned Lean library, inspects theorem statements and axioms, regenerates the selected large certificates from portable plans, and evaluates those same representations. Upstream regeneration reconstructs the diagnostic plans from the identified frozen provider values; it has additional input requirements. The native physics path builds the locked arTeMiDe/LHAPDF stack and reproduces the external comparisons. Formal replay does not require native LHAPDF or arTeMiDe.

The proof project is independently pinned to Lean 4.21.0 with no mathlib dependency. The Python numerical workflow was recorded in project-local environments; the external benchmark used Python 3.11 with NumPy 1.26.4 and SciPy 1.14.1, serial GNU Fortran, and LHAPDF 6.5.5. The archive records exact build, source, model, and grid identities. Distinct public test scopes distinguish ordinary runtime checks from fresh kernel checking. Portable finite-plan fixtures allow the formal example to be reproduced without a large provider archive or knowledge of the development handoffs.

\section{Discussion and conclusions}
\label{sec:conclusion}
The method has two complementary results. In the external calculation, an observation error that survives dimensional and array-shape checks produces a substantial evolution-parameter displacement with only a moderate residual. The comparison contract rejects the unsupported operation before fitting, and the admitted bin map restores the intended synthetic extraction. At the representation layer, the concrete Lean checker establishes preservation under dependency-safe caching for all admitted parameter environments. The categorical composition theorem specifies how a later cache reference must be transported when checked stages are assembled.

The advantage over an isolated equality unit test is therefore not a claim that conventional testing is incapable of finding the illustrated errors. It is that observation meanings, parameter dependencies, candidate transformations, and evidence become explicit and reusable elements of the calculation. The ordinary numerical test remains essential for the binding and approximation layers. The formal result protects a supported class of transformations universally, while the external example shows why correctly identifying an observation matters to an actual physics workflow.

The two guarantees are connected through the finite source, but their boundaries remain distinct. The observation constructor is runtime-validated, and the finite kernel-geometry certificate uses an analytic primitive binding. A source-to-plan proof does not establish the physical correctness of the source or native backend. Likewise, certifying chart A and chart B separately does not prove their physical equivalence; that comparison retains its analytical definition and numerical tests. This separation permits a useful formal statement without disguising unproved decoder, numerical, or physical assumptions.

The categorical structure is deliberately concrete: fixed signatures, parameter interfaces, admissible substitutions, and an interpretation into functions. It supplies local-to-global reuse through checked staged composition. A more general category of hadronic operators, schemes, and process families would require additional physical maps and domain conditions. Yoneda provides a general relationship-based characterization once such a category is supplied~\cite{Riehl2016}; the finite observation vector in this work is not that complete representable profile, and no general TMD uniqueness result follows from the finite data.

The present validation also remains bounded. The external study has one free parameter, frozen other inputs, and synthetic data on published geometry. The supporting two-parameter model uses a finite Fourier window and a deliberately limited coefficient inventory. Its coverage check is conditional on one truth and covariance, with finite Monte Carlo precision. The reported runtime experiment measures the observation adapter over the same backend, not the cost of a fully verified extraction. These scopes identify what the released evidence supports and which extensions need new work.

Several extensions can reuse the architecture. Exact chart substitutions, verified common-subexpression sharing, finite-sum reindexing, additional observation operations, and quantitative error propagation require new supported rules rather than informal relabeling. More flexible nonperturbative families and coupled datasets require corresponding identifiability and model-dependence studies. An AI system can propose an expression or evaluation plan while an independent checker enforces established rules, retaining rejected proposals and unresolved obligations. Broader constructive or homotopical factorization programs additionally need physical realization and remainder theorems; the current framework supplies a practical representation and verification starting point, not those future results.

In summary, reproducibility of a TMD extraction benefits from preserving both what is observed and what function is executed. The external observation experiment demonstrates the first requirement, and the concrete verified plan language establishes a reusable part of the second. Their combination yields a computational method with explicit scientific scope, executable failure diagnostics, and a formally checked preservation guarantee.

\appendix
\section{Diagnostic TMD, evolution, and boundary transport}
\label{app:diagnostic}
This appendix specifies the supporting finite-window calculation of Sec.~\ref{sec:validation}; it is not the external SV19 prescription. Its context declares a Collins soft-subtracted unpolarized quark TMD with $\overline{\mathrm{MS}}$ ultraviolet subtraction and the past-pointing Drell--Yan Wilson-line character. The numerical realization begins from a declared boundary family rather than an evaluated regulated operator matrix element.

Let
\begin{equation}
 b_*=\frac{b}{\sqrt{1+(b/b_{\max})^2}},\quad
 \mu_b=\frac{2e^{-\gamma_E}}{b_*},\quad b_{\max}=0.5\GeV^{-1}.
 \label{eq:diagprofile}
\end{equation}
The boundary at $s_b=(\mu_b,\mu_b^2)$ is
\begin{equation}
 F_{q/H}(x,b;s_b)=f_{q/H}(x,\mu_b)e^{-ab^2}.
 \label{eq:diagboundary}
\end{equation}
The provider returns $f$, with an explicit division by $x$ when importing an LHAPDF \softcode{xfxQ} value. With $c(\mu)=C_F\alpha_s(\mu)/\pi$ and $C_F=4/3$, use
\begin{align}
 \frac{\partial F}{\partial\ln\mu}&=\gamma_F^{(\lambda)}(\mu,\zeta)F,
 &\frac{\partial F}{\partial\ln\sqrt\zeta}&=K(b;\mu)F,\nonumber\\
 \gamma_F^{(\lambda)}&=c(\mu)\left[\frac32\lambda-\ln\frac{\zeta}{\mu^2}\right],
 &\gamma_K&=2c(\mu),
 \label{eq:diaggen}\\
 K(b;\mu)&=-g_2b^2-2\int_{\mu_b}^{\mu}c(\nu)\,\dd\ln\nu.
 \label{eq:diagK}
\end{align}
This completion obeys $\partial_{\ln\mu}K=\partial_{\ln\sqrt\zeta}\gamma_F^{(\lambda)}=-\gamma_K$. It defines an integrable scalar connection on the admitted positive-scale domain. The choice of completion and the distinction between truncated ingredients and path-independent evolution matter in TMD calculations~\cite{ScimemiVladimirov2018}. Scale integrals are oriented; no cap $\mu_b\le Q$ is imposed.

At $s_Q=(Q,Q^2)$ define
\begin{align}
 P_\lambda(b,Q)&=\int_{\mu_b}^{Q}c(\mu)
  \left[\frac32\lambda-2\ln\frac Q\mu\right]\dd\ln\mu,
 \label{eq:P}\\
 F_{q/H}(x,b;s_Q)&=f_{q/H}(x,\mu_b)\exp\!\left[P_\lambda(b,Q)\right.\nonumber\\
 &\hspace{8pt}\left.{}-b^2\left(a+g_2\ln\frac Q{\mu_b}\right)\right].
 \label{eq:canonical}
\end{align}
Writing multiplicative evolution and carrying signed inputs directly avoids taking logarithms of PDF values. The baseline has $\lambda=1$; the control uses $\lambda=0.8$. Both satisfy the integrability relation, but their coefficient definitions differ.

The two fixed reference scales are $s_A=(5\GeV,25\GeV^2)$ and $s_B=(10\GeV,100\GeV^2)$. The full B boundary in coordinates $\beta=a+g_2\ln2$ is
\begin{equation}
\begin{split}
 F^B_{q/H}(x,b;s_B)={}&f_{q/H}(x,\mu_b)\exp\!\left[P_\lambda(b,10\GeV)\right.\\[-2pt]
 &\left.{}-b^2\left(\beta+g_2\ln\frac{5\GeV}{\mu_b}\right)\right].
\end{split}
\label{eq:Bboundary}
\end{equation}
Thus $\beta$ is not the complete width of a freshly imposed Gaussian at $s_B$. The image domain obeys $0\le g_2/\mathrm{GeV}^2\le0.10$ and $0.10\le(\beta-g_2\ln2)/\mathrm{GeV}^2\le0.50$. These complete families yield Eq.~\eqref{eq:argmin}. An affine covariance transport $C_B=T C_A T^{\mathsf T}$, with $T=\left(\begin{smallmatrix}1&\ln2\\0&1\end{smallmatrix}\right)$, is exact as a coordinate relation; it does not make an inverse-Fisher approximation exact.

Only two scale moments are needed,
\begin{equation}
 M_k(\mu)=\int_{\mu_\dagger}^{\mu}c(\nu)
   \left[\ln\frac\nu{\mu_{\rm ref}}\right]^k\dd\ln\nu,
 \qquad k=0,1,
 \label{eq:moments}
\end{equation}
where $\mu_{\rm ref}=1\GeV$ and the common lower reference cancels in differences. These moments supply both the canonical expression and supported rectangular transport paths. The fixed query plan contains 156,928 collinear-density values and 316,663 coupling values, exported with canonical binary64 coordinates. Native replay adds no new interpolation or nearest-point substitution; the native provider's own interpolation remains part of the fixed input.

Production uses 32-point Gauss--Legendre integration on the declared $b$ and logarithmic-scale panels; independent refinements use order 16 in one direction at a time. The reference uses composite two-point Gauss rules with 64 subpanels in $b$ and eight in logarithmic scale, compared with 32 and four. Native threshold structure is included in the panel definition where available. Numerical acceptance covers parameter corners and selected fit points rather than only the generating parameters.

For the two-leg integrand, the parameter derivatives multiply the integrand by $-2b^2$ and $-2b^2\ln(Q/\mu_b)$. Their independent derivative and integration checks support the sensitivity calculation. The coefficient control changes the two-leg exponent by
\begin{equation}
 3(\lambda-1)\int_{\mu_b}^{Q}c(\mu)\,\dd\ln\mu.
 \label{eq:lambda-change}
\end{equation}
Constant shifts in $a,g_2$ cannot absorb this as an identity on a nondegenerate $b$ interval at two distinct hard scales: subtracting the two required identities leaves a term proportional to $b^2\delta g_2\ln(Q_2/Q_1)$ equated to a nonzero $b$-independent integral. The finite Fourier observations still require a numerical check of distinguishability, supplied by the recorded refit.

\section{Formal declarations and certified scope}
\label{app:inventory}
The proof library is located in \softcode{formal/007/TypedTMDPlan}. Its versioned directory name is a software path, not a prerequisite for understanding the paper. The two principal files are \softcode{Core.lean} and \softcode{Category.lean}; \softcode{Examples.lean} supplies concrete examples. Table~\ref{tab:inventory} groups all 24 audited declarations. Names in the category group have the prefix \softcode{TypedTMDPlan.Category.Arrow}; the other names have prefix \softcode{TypedTMDPlan}. An inventory records the actual statements and axioms for each, including helper results.

\begin{table*}[tbp]
\caption{Complete declaration groups in the reviewed Lean inventory. The groups contain 8, 11, and 5 declarations, respectively. Their count is an audit aid, not a measure of mathematical novelty.}
\label{tab:inventory}
\small
\begin{tabularx}{\textwidth}{@{}p{.18\textwidth}Y@{}}
\toprule
Group & Declarations\\
\midrule
Category and interpretation & \softcode{ext}, \softcode{id\_comp}, \softcode{comp\_id}, \softcode{comp\_assoc}, \softcode{interpret\_id}, \softcode{interpret\_comp}, \softcode{fixed\_slice\_preserved}, \softcode{checked\_composition}.\\
Core syntax and validator & \softcode{congrTwo}; \softcode{Expr.dependency\_sound}, \softcode{Expr.interpretation\_unique}, \softcode{Expr.subst\_vars}, \softcode{Expr.subst\_comp}, \softcode{Expr.eval\_subst}; \softcode{cacheable\_sound}, \softcode{cacheable\_subst}, \softcode{Plan.safe\_sound}, \softcode{check\_sound}, \softcode{live\_accepted}.\\
Examples & \softcode{Examples.good\_accepts}, \softcode{Examples.bad\_rejects}, \softcode{Examples.good\_preserves}, \softcode{Examples.fixed\_g\_cache\_is\_admissible}, \softcode{Examples.quadrature\_accepts}.\\
\bottomrule
\end{tabularx}
\end{table*}

The source/plan equality tested by the validator preserves primitive order and source erasure. It does not use algebraic normalization. The generic theorem quantifies over a fixed interpretation on an arbitrary value type, and the formal language is single-sorted at the value level. Physical units, process labels, measured densities, and high-level source lowering remain runtime/source obligations. Proof fields of admitted arrows establish the computed fixed-coordinate condition, not a postulated equality of arbitrary responses.

The positive certificate set comprises the four cases of Table~\ref{tab:certificates}; the negative set contains the explicitly frozen fitted-kernel candidate. Each certificate binds literal source and plan definitions to a recorded wire identity. Rechecking a supplied and a regenerated snapshot provides provenance for both numerical realizations without assuming their upstream coefficients are bitwise identical across platforms. Exact preservation of each fixed plan and cross-platform agreement of its independently produced coefficients are distinct statements.

The historical abstract Lean seed is retained in the project as an earlier structural example. It is not the evidence for Theorems~\ref{thm:check} and~\ref{thm:composition}, which concern the concrete dependency analysis, checker, and category developed here.

\section{Numerical discrepancy in the inference metric}
\label{app:error}
For a positive-definite fixed covariance, let $r=\Sigma^{-1/2}(\mu-d)$ and let $e=\Sigma^{-1/2}\delta\mu$ denote a numerical response change. Expansion and Cauchy--Schwarz give
\begin{equation}
 \left|\|r+e\|_2^2-\|r\|_2^2\right|
 \le2\|r\|_2\|e\|_2+\|e\|_2^2.
 \label{eq:objective-error}
\end{equation}
This exact inequality translates an actual response-error bound into an objective-error bound. In the executed numerical checks, $e$ is estimated from independent comparisons and refinements. The resulting budgets are consequently empirical diagnostics, not certified bounds on a continuum integral. The two-route budgets reported in Sec.~\ref{sec:external} sum the respective contributions for the compared predictions.

For the coverage diagnostic, an orthogonal projector of rank two applied to a standard Gaussian has a chi-squared squared norm with two degrees of freedom. Parameter correlation changes coordinate geometry but not the projector's rank. The bounded linear calculation solves the corresponding two-dimensional constrained quadratic on the same parameter box, and its numerical solutions were cross-checked by an independent bounded linear least-squares routine. The nonlinear calculation uses the frozen original fitted results. Thus the paired diagnosis changes the model geometry being analyzed while holding the random fluctuations fixed; it neither resamples until a favorable outcome nor changes the likelihood thresholds.

A nonzero smallest singular value supports local sensitivity in the tested finite-dimensional family. It does not prove global identifiability of a general TMD. Similarly, an exact preservation theorem for a computational plan does not establish the adequacy of the fitting family or uncertainty model. These distinctions remain part of the stated comparison contract.

\section{Source identities and reproduction entry points}
\label{app:reproduction}
The external engine is fixed to arTeMiDe v2.05, commit
\begin{lstlisting}
ea0af1a75e21e316c1ac4ece51933988836a6650
\end{lstlisting}
and its data-definition source is the DataProcessor commit
\begin{lstlisting}
18967d1c52dabf93e17a640405b548071904cfcc
\end{lstlisting}
The selected input paths are \path{Models/SV19/constants-files/DY_nnlo/const-NNPDF31_NNLO}, \path{Models/SV19/Replicas/DY_nnlo/DY_NNPDF31_nnlo.rep}, and \path{DataLib/unpolDY/A8-00y04.csv}. The nine fixed-vector entries before varying $c_0$ are
\begin{equation}
\begin{split}
 (&1.86041,\ 0.029551,\ 0.253434,\ 9.04351,\ 346.999,\\[-2pt]
  &2.47992,\ -5.69988,\ 0,\ 0).
\end{split}
\label{eq:npvector}
\end{equation}
Their mixed units and meanings follow the pinned source; they are not interchangeable with the diagnostic model's parameters. The PDF set is \softcode{NNPDF31\_nnlo\_as\_0118\_1000}, member 0. The source lock includes hashes of model files, configuration, selected row, dataset geometry, grids, and precision changes, and distinguishes external licenses from the original MIT code.

A source-checkout route begins with the exact public version:
\begin{lstlisting}
git clone https://github.com/zetanaut/typed-TMD.git
cd typed-TMD
git checkout v0.4.0
python3.11 -m venv .venv
. .venv/bin/activate
python -m pip install -e .
PYTHONPATH=src python -m unittest discover -s tests -v
\end{lstlisting}
A library build is separate from literal-plan acceptance:
\begin{lstlisting}
cd formal/007
lake build
\end{lstlisting}
The release documentation supplies the corresponding certificate and replay entry points and their input manifests. Full formal reproduction must include the selected large literal certificates, negative case, theorem/axiom inventory, and evaluations of the same accepted plans, not merely a successful library build. The retained large-certificate stack setting is recorded in Sec.~\ref{sec:certificate}. Native benchmark instructions have additional source and library requirements; they are not prerequisites for finite-plan replay.

The manuscript's figure scripts consume a compact projection of archived result fields with parent hashes. They recompute no fit and make no new PDF or native-kernel query. Full precision values, failed numerical attempts, and accepted execution records remain in the software/evidence archive. This article reports rounded values appropriate to their interpretation, while version-specific artifacts support more detailed comparisons.

\section*{Data and software availability}
The original implementation, Lean proof sources, portable finite plans, example scripts, and reproduction documentation are released as typed-TMD v0.4.0~\cite{TypedTMD2026}, under the version DOI \href{https://doi.org/10.5281/zenodo.23546800}{10.5281/zenodo.23546800}. Synthetic observations and saved numerical evidence are identified by the accompanying manifests. External software and PDF inputs are retrieved from their pinned sources according to their own redistribution terms. The ATLAS source is used for published bin geometry and conventions, not as a newly fitted experimental dataset.


\bibliography{references}
\end{document}